\documentclass[aps,prd,twocolumn,showpacs]{revtex4}
\usepackage[english]{babel}
\usepackage{amsmath,amssymb}
\usepackage{graphicx}

\newcommand{\be}{\begin{equation}}
\newcommand{\ee}{\end{equation}}
\newcommand{\bq}{\begin{eqnarray}}
\newcommand{\eq}{\end{eqnarray}}
\newcommand{\snf}{\sin(\bar \mu c)}
\newcommand{\snfs}{\sin^2(\bar \mu c)}
\newcommand{\csf}{\cos(\bar \mu c)}

\newcommand{\heff}{{\cal H}_{\mathrm{eff}}}

\newcommand{\hcl}{{\cal H}_{\mathrm{cl}}}

\newcommand{\hm}{{\cal H}_{\mathrm{M}}}
\newcommand{\hme}{{\cal H}_{\mathrm{M eff}}}

\newcommand{\rcr}{\rho_{\mathrm{crit}}}

\newcommand{\f}{\frac}

\newcommand{\mubar}{\bar{\mu}}

\newcommand{\half}{\frac{1}{2}}

\newcommand{\Pp}{\Pi_\phi}

\begin{document}


\title{Non-Singular Bouncing Universes in Loop Quantum Cosmology}
\author{Parampreet Singh\footnote{e-mail address: {\tt singh@gravity.psu.edu}}, Kevin Vandersloot\footnote{e-mail address: {\tt kfvander@gravity.psu.edu}}}
\affiliation{Institute for Gravitational Physics and Geometry, The Pennsylvania State University,
University Park, PA 16802, USA}
\author{G. V. Vereshchagin\footnote{e-mail address: {\tt veresh@icra.it}}}
\affiliation{ICRANet and ICRA, P.le della Repubblica 10, 65100 Pescara, Italy, \\
University of Rome "La Sapienza", Physics Department, \\
P.le A. Moro 5, 00185 Rome, Italy.}

\begin{abstract}
Non-perturbative quantum geometric effects in Loop Quantum Cosmology
predict a $\rho^2$ modification to the
Friedmann equation at high energies. The quadratic term is negative
definite and can lead to generic bounces when the matter energy density becomes equal to a critical value of the order of the Planck density. The non-singular bounce is
achieved for arbitrary matter without violation of positive energy
conditions. By performing a
qualitative analysis we explore the nature of the bounce for inflationary
and Cyclic model potentials. For the former we show that inflationary
trajectories are attractors of the dynamics after the bounce
implying that inflation can be harmoniously embedded in LQC. For the latter difficulties
associated with singularities in cyclic models can be overcome. We
show that non-singular
cyclic models can be constructed with a small variation in the
original Cyclic model potential by making it slightly positive in the
regime where scalar field is negative.
\end{abstract}

%



\maketitle

\section{Introduction}


One of the cornerstone problems in cosmology is that of the initial conditions of our universe. In standard big bang
cosmology the weight of this problem is shifted to quantum gravity which is expected to generate ideal conditions for 
the genesis of our universe. Generally, it is envisioned that our universe started in a hot expanding phase 
from the big bang and entered a stage of inflation before becoming radiation and matter dominated. However the idea that our expanding universe was preceded by a contracting phase of a classical
universe has had many avatars. This idea, though promising to eliminate some of the fundamental problems related with 
initial conditions in standard cosmology and providing an alternative to inflation, has remained unsuccessful
due to absence of  generic non-singular evolution from the contracting
to the expanding branch. For a flat $k = 0$ cosmological model, which
is of interest of the present work, based on classical general
relativity (GR) such an evolution is forbidden by  powerful singularity
theorems, unless one assumes an exotic form of matter which violates
positive energy conditions. Though quantum corrections have been
introduced to achieve a non-singular evolution between contracting and
expanding phases in special conditions, it has been realized that a
generic non-singular evolution is difficult to achieve unless one incorporates
non-perturbative effects of quantum gravity.
In absence of any non-perturbative quantum gravitational
modifications to dynamics, innovative ideas like pre big bang
string cosmology \cite{pbb} and Ekpyrotic/Cyclic models \cite{ekp-cyclic,StTur} 
have so far had limited viability \cite{cyc-nonsing}.

Loop quantum gravity (LQG) is a leading non-perturbative background
independent approach to quantize gravity \cite{lqg_review}. The underlying geometry in
LQG is discrete and the continuum spacetime is obtained from quantum
geometry in a large eigenvalue limit. The application of LQG techniques to
homogeneous spacetimes results in loop quantum cosmology
(LQC) \cite{mblr} which has led to important insights on the resolution of
singularities in various situations \cite{bigbang,abl,sing}.
Using extensive analytical and numerical methods the analysis
of the evolution of semi-classical states for the flat $k=0$ model
with a massless scalar field has shown that a contracting
semi-classical universe passes through the quantum regime
bouncing to an expanding semi-classical universe \cite{aps,aps1,aps2}.
Thus within the context of the model considered the idea of
the non-singular bounce is realized in a natural fashion.
 

The underlying dynamics in LQC is governed by a discrete quantum
difference equation in quantum geometry. However, using semi-classical states we can
construct an effective Hamiltonian description on a {\em continuum}
spacetime which has been shown to very well approximate the quantum
dynamics \cite{sv,aps1,aps2}. In particular, we can obtain effective equations for the modified Friedmann
dynamics which can be used to investigate the role of
non-perturbative quantum corrections. 
An important feature of the dynamics is that
the classical Friedmann equation is modified to include a $\rho^2$
term which is relevant in the high energy regime. The 
modified term in the Friedmann equation is negative definite 
implying a bounce (or more generally a turn-around) 
when the energy density reaches a 
critical value on the order of the Planck density in accordance with
the results from the quantum evolution. 

An important question which gains immediate attention is the viability of the bounce picture for more general matter 
sources and compatibility of inflationary and Cyclic model ideas with LQC. In this regard various questions arise: Does the
bounce occur for various inflationary potentials? 
Do inflationary solutions 
form the attractors of the dynamics after the bounce?  Since the modifications which lead to a non-singular bounce 
are obtained from non-perturbative quantum gravity effects is it possible that these can lead to construction of non-singular cyclic models using Cyclic/Bicyclic potentials? What insights are obtained to alleviate the problems of Cyclic models? 

This work seeks to  answer  these questions using the effective theory in LQC.
The plan of this paper is as follows. In the next section we briefly review the effective theory of LQC and classify the differences between quantum bounce generated by loop quantum dynamics and the classical bounce originating from 
general relativity. In Sec. III we perform a qualitative analysis 
using phase space diagrams to study
the dynamics for inflationary potentials ($\phi^2$
and $\phi^4$) and show that inflationary trajectories are attractors of the dynamics after the bounce. In Sec. IV we apply 
the qualitative analysis to negative potentials of the Cyclic/Bicyclic models. We show that the Bicyclic potential \cite{bike}
can successfully lead to non-singular cycles, whereas the original Cyclic potential 
can exhibit cycling of the scale factor but not the scalar field. A truly cyclic model can be easily constructed if the Cyclic potential becomes 
slightly positive in the regime where the scalar field is negative, as the Bicyclic potential does. The non-singular cyclic behavior obtained for negative potentials is unavoidable and occurs for generic values of parameters. We conclude the paper with a discussion and open issues in Sec V.

\section{Effective dynamics  in loop quantum cosmology}

LQC is a canonical quantization of homogeneous spacetimes based on
techniques used in LQG. As in LQG, the classical phase space for the gravitational
sector is denoted by two conjugate variables the connection and the
triad, which encode
curvature and spatial geometry respectively.
Using symmetries of the homogeneous and isotropic
spacetime the dynamical part of the connection is determined by a single
quantity labeled $c$ and likewise the triad is determined by a
parameter $p$. 
For the flat $k=0$ model (which we will only consider in this work)
on classical solutions of general relativity
the relations between the new variables and the usual
metric variables is
\be \label{c,p}
	c = \gamma \, \dot{a} ~, ~~~
	p = a^2 
\ee
where $\gamma$ is the Immirzi
parameter whose value $\gamma \approx 0.2375$ is typically constrained by black hole entropy
considerations \cite{bek_hawking}.

In the Hamiltonian formulation for homogeneous and isotropic space-times the dynamical 
equations of motion can be determined completely
by the Hamiltonian constraint. Under quantization, the Hamiltonian
constraint gets promoted to an operator and the quantum wave-functions
are annihilated by that operator. It is thus at the level of the
Hamiltonian constraint that modifications due to LQC will appear and from
the modified Hamiltonian constraint, the effective Friedmann constraint
will be derived. Classically in terms of the connection-triads variables
the Hamiltonian constraint is given by
\be \label{Hclass}
	\hcl = - \frac{3 \sqrt{p}}{\kappa \gamma^2} \, c^2 + \hm
\ee
with $\kappa = 8 \pi G$ (where $G$ is Newton's gravitational constant)
and $\hm$ being the matter Hamiltonian. The
complete equations of motion are derived from  Hamilton's
equations $\dot{x} = \{ x , \hcl \}$ for any phase space variable $x$, and by enforcing that
$\hcl$ should vanish. The variables $c$ and $p$ are canonically conjugate
with Poisson bracket $\{c,p\} = \gamma \kappa / 3$ the use of which
gives us relation $c = \gamma \dot a$ from Hamilton's equation for  $\dot{p}$ together
with $p = a^2$. Substituting these relations  into the Hamiltonian
constraint (\ref{Hclass}) and enforcing that the constraint vanishes then gives
the classical Friedmann equation.

The elementary variables used for quantization in LQC are the triads
and holonomies
of the connection: $h_i(\mu) = \exp (\mu c
\tau_i) = \cos(\mu c / 2) + 2 \sin(\mu c / 2 )\, \tau_i$. Here
$\tau_i$ is related to Pauli spin matrices as $\tau_i
= - i \sigma_i/2$ 
and $\mu$ is the eigenvalue of the triad operator $\hat p$. The
gravitational constraint is expressed in terms of these elementary
variables and quantized. 
 On quantization the Hamiltonian constraint leads to a
discrete quantum difference equation whose solutions are
non-singular \cite{bigbang,abl}.

The analysis of the quantum Hamiltonian using semi-classical
states belonging to the physical Hilbert space reveals that on
backward evolution of our expanding phase of the universe, the
universe bounces at a critical density into a contracting branch. By
finding expectation values of the Dirac observables we can investigate
the differences between quantum and classical dynamics. It turns out
that quantum geometric effects become dominant only when the energy
density $(\rho)$ of the universe becomes of the order of the critical density 
$\rcr$ \cite{aps2,abhay} and classical general
relativity is a good approximation to the quantum dynamics when $\rho
\ll \rcr$.

The quantum dynamics of LQC can be well approximated by an effective
description. Several proposals have been made to derive this
effective theory \cite{josh,vt,sv,dh} and the result is
that the dynamics can be described in terms of an
effective Hamiltonian constraint which to leading order is \cite{eff_ham_fn}

\be
\heff = - \f{3}{\kappa \gamma^2 \bar \mu^2}  \,a \sin^2(\bar \mu c) +
\hm ~.
\ee
where $\bar \mu = \sqrt{3\sqrt{3}/{2|\mu|}}$ \cite{aps2} and $\mu$ is the
eigenvalue of the triad operator $\hat p$ in the quantum theory
\cite{abl}:
\be
\hat p \, |\mu\rangle = \f{8 \pi \gamma l_{pl}^2}{6} \, \mu \,
|\mu\rangle ~. \label{jueq}
\ee
Here $l_{pl} = \sqrt{\hbar G}$ denotes the Planck length (in units of
speed of light equal to unity).
In this work we will be
mainly interested in the matter Hamiltonians corresponding to a
massive scalar field $\phi$ with momentum $\Pi_\phi$ and potential
$V(\phi)$:
\be \label{hmatter}
\hm = \frac{1}{2} \frac{\Pi_\phi^2}{p^{3/2}} + p^{3/2} \, V(\phi) ~.
\ee
The energy density and pressure for the scalar field
are given by the same expressions as those classically, i.e.
\be \label{edp}
\rho = \rho_\phi = \frac{1}{2} \, \dot \phi^2 + V(\phi), ~~~  p_\phi = \frac{1}{2} \, \dot \phi^2 - V(\phi) ~
\ee
where we have used Hamilton's equations to get $\dot \phi =
\Pi_\phi/p^{3/2}$. 
It is then straightforward to find the Klein-Gordon equation using
Hamilton's equations for $\dot \phi$ and $\dot \Pi_\phi$, which turns out to be of the same form as the classical equation
and therefore the scalar field satisfies the stress-energy conservation law:
\begin{equation}\label{EC} 
\dot \rho_\phi + 3 \frac{\dot a}{a} (\rho_\phi + p_\phi) = 0~.
\end{equation}

The modified Friedmann equation can be found by using  Hamilton's
equations for $\dot p$
\be
\dot p = \{p,\heff\} = - \f{\gamma \kappa}{3} \f{\partial}{\partial c}\heff
 = \frac{2 a}{\gamma \bar \mu} \, \snf \, \csf  \label{dotp}
\ee
which on using Eq.(\ref{c,p}) implies that the rate of change of the scale factor
is given by
\be
\dot a = \frac{1}{\gamma \bar \mu} \, \snf \, \csf ~. \label{dota} 
\ee
Furthermore, the vanishing of the Hamiltonian constraint implies
\be \label{sin}
\snfs = \frac{\kappa \gamma^2 \bar \mu^2}{3  \, a} \, \hm
\ee
Squaring equation (\ref{dota}) and plugging in (\ref{sin}) yields the effective Friedmann equation
for the Hubble rate $H=\dot{a}/a$
\be 
H^2 = \frac{\kappa}{3} \, \rho \left( 1 - \frac{\rho}{\rcr} \right) \label{FE1} ~.
\ee 
with the critical density given by 
\be
\rcr = \frac{\sqrt{3}}{16 \pi^2 \gamma^3}\rho_{pl} \label{rcr}
\ee
where $\rho_{pl} = 1/(\hbar G^2)$ is the Planck density. 
The non-perturbative quantum geometric effects are manifested in
the form of a $\rho^2$ modification of the Friedmann equation. Since
the modified term is negative definite, when $\rho = \rcr$ the Hubble
parameter vanishes and the universe experiences
a turn-around in the
scale factor. For $\rho \ll\rcr$, the  modifications to Friedmann dynamics become negligible and we recover standard  Friedmann dynamics. Note that origin of $\rcr$ is purely quantum, since  as $\rho_{pl} \propto 1/\hbar$, $\rcr \rightarrow \infty$ in the classical limit $\hbar \rightarrow 0$.

Interestingly, similar $\rho^2$ modifications have also been
obtained in Randall-Sundrum braneworld scenarios based on string
theory but those come
with a positive sign in the Friedmann equation and a bounce is not
possible.  If one assumes the
existence of a time-like extra dimension then the sign can become
negative as in LQC \cite{sahni} (for an analysis of dynamics of the
brane-world model with two time-like dimensions see Ref. \cite{piao}
and  \cite{jldm}).
A detailed
comparison of the effective dynamics of LQC and braneworld scenarios
shows interesting features which include a dual relationship between
their scaling solutions \cite{singh:2006a}. 

\section{Quantum and Classical bounces}
Having established the equations for the effective Friedmann dynamics
we now wish to classify the conditions under which effective dynamics given
by the modified Friedmann equation (\ref{FE1}) implies a turn-around; namely
either a bounce or re-collapse. As discussed above when the energy density
of the universe reaches the critical value $\rcr$, the Hubble
parameter becomes zero and a bounce/re-collapse is expected.
Since we will also consider the possibility of the scalar field potential
being negative we also discuss the case where the matter energy
density vanishes which also leads to a turn-around.

More precisely from Eq. (\ref{dota}) we can determine the conditions under which such
a turn-around
can occur; i.e., when $\dot{a} = 0$.
It is clear from that equation that there are two possibilities, namely
when $\mubar c = (n+\half)\pi$ and $\mubar c = m \pi$ for any  integers $n,m$.
The energy density can be computed for both cases and is equal to
$\rcr$ for the former and zero for the latter. We thus
label turn-arounds occurring due quantum geometric effects when $\mubar
c = (n+\half)\pi$ as {\it quantum turn-arounds} and those which occur
when energy density vanishes for $\mubar c = m \pi$ as {\it classical turn-arounds}.

To determine whether or not a bounce or recollapse occurs we need to 
go further and calculate
the second time derivative of the triad (which is proportional to $\ddot{a}$ at
the turn-around), where a negative value indicates
a recollapse and a positive value a bounce. 
Taking the time derivative of Eq. (\ref{dotp}) and evaluating
when $\dot{p}$ equals zero gives
\begin{equation}
        \ddot{p}\, |_{\dot{p}=0} = \frac{2 \sqrt{p}}{\gamma} \cos(2 \mubar c) \dot{c} \,.
\end{equation}
To determine the sign of $\ddot{p}$ we therefore need to know
the time derivative of $c$ which is determined from Hamilton's equations
 $\dot{c} = \{c, \heff \}$ leading to 
\begin{eqnarray}
        \dot{c} &=& \nonumber\frac{c}{\gamma^{3/2} \sqrt{2\pi \sqrt{3}}\, l_{pl}} \sin(\mubar c)
\cos(\mubar c) - \nonumber \\ & &
        \frac{3 \sqrt{p}}{4 \pi \sqrt{3} \gamma^2  l_{pl}^2} \sin^2(\mubar c) + \frac{\gamma \kappa}{3} \frac{\partial \hm}{\partial p} = \nonumber \\ & & - \frac{3 \sqrt{p}}{4 \pi \sqrt{3} \gamma^2   l_{pl}^2} \sin^2(\mubar c) + \frac{1}{3}\gamma \kappa \frac{\partial \hm}{\partial p} ~.
\label{paccel}
\end{eqnarray}
where the first term vanishes for both quantum and classical turn-arounds.
 We now need to consider separately the quantum and classical
turn-arounds and to specify the form of matter. We first consider
the quantum turn-arounds.

\subsection{Quantum Turn-arounds}
The quantum turn-arounds as stated above are a direct consequence of the discrete quantum
geometry effects of LQC. 
From the fact that $\mubar c = (n+\half)\pi$ we
have $\sin(\mubar c) =1$ and $\cos(\mubar c)=0$ which implies 
the matter energy density satisfies $\rho =
\rcr$. Under these conditions the equation (\ref{paccel}) for the triad acceleration becomes 
\begin{equation}
\ddot{p}\, |_{\dot{p}=0} = \kappa p
                \left( \rcr - \frac{2}{3} \frac{1}{\sqrt{p}} \frac{\partial
\hm}{\partial p} \right)
\end{equation}
                                                                                
Let us first consider the simplest case of matter with constant
equation of
state $w$ with energy density $\rho \propto a^{-3(1+w)} =
p^{-3/2(1+w)}$ which implies
 $\hm \propto p^{-3w/2}$. It is then straightforward to calculate
$\ddot{p}$
giving
\begin{equation}
        \ddot{p}\, |_{\dot{p}=0} = \kappa p \rcr (1+w)
\end{equation}
and thus we can make the identification that for $w<-1$ a recollapse can occur
and for $w>-1$ a bounce can occur. 
This can be understood in a simple fashion. Matter with
equation of state $w<-1$ violates the null energy condition
and hence has increasing energy density as the universe expands.
Thus when the energy density hits the quantum critical value
a re-collapse occurs and the universe begins contracting.
 In the classical
universe the energy density of such  matter can grow to infinite
values and even  lead to a 
singularity (for example a big rip singularity encountered for phantom
field which has a negative kinetic energy \cite{phantom}). In the context
of LQC the implication is that such a singularity does not occur
as the energy density remains bounded and the big rip is avoided \cite{phantom1}.
On the other hand matter with $w > -1 $ has increasing energy
density as the universe contracts and therefore a quantum bounce
will occur when the critical density is reached.

Let us now turn to the  case of a massive scalar field. Here we
consider
general forms of the potential which can be negative valued and phantom models
where the kinetic term is negative definite. The general form of the matter
Hamiltonian is
\begin{equation}
        \hm = \pm \half p^{-3/2} \Pp^2 + p^{3/2} V(\phi)
\end{equation}
where the $\pm$ indicates normal and phantom matter respectively.
With this form we
can calculate $\frac{\partial \hm}{\partial p}$ which gives
\begin{equation}
        \frac{\partial \hm}{\partial p} = \frac{3}{2p} \left(
                \mp \half p^{-3/2} \Pp^2 + p^{3/2} V(\phi) \right)
\end{equation}
and we can simplify this by noting that when $\rho=\rcr$ we have
$\pm\half p^{-3/2} \Pp^2 = p^{3/2} (-V + \rcr)$ whence $\ddot{p}$ simplifies
to
\begin{equation}
        \ddot{p}\, |_{\dot{p}=0} = 2 \kappa p \left( \rcr - V \right) \,.
\end{equation}
The implication of this is that if the scalar field potential is
less that the critical density at the turn-around then
a bounce occurs, otherwise if the potential is larger a recollapse
occurs. In this form we can analyze the turn-around behavior for the different
types of scalar field and potential.

We start with a normal scalar field
where the kinetic term of the matter energy density is
positive definite. Because of this when the total energy
density is equal to the critical value at the turn-around the potential {\em must}
be less than than $\rcr$. Thus for a normal scalar field, only a bounce can
occur
in a quantum turn-around. This can be understood intuitively from
the previous paragraph given that
 an ordinary scalar field can not have $w<-1$. Since we have not made any assumptions on the form of the
potential 
even negative potentials will not lead to a quantum recollapse (though
as we shall see a negative potential can lead to a classical recollapse).

Considering now phantom models with a negative definite kinetic term we find
the
opposite situation. First a positive potential is required in general
from the effective Friedmann equation to get classical behavior of the scale
factor.
Furthermore since the kinetic term is negative definite 
when the matter density equals the critical density, the potential is larger
than the critical density. Hence $\ddot{p} < 0$ and the phantom field
behaves like a matter with $w<-1$ and a quantum recollapse can occur. This implies
generically that a phantom field can not have a quantum bounce in LQC.
                                                                                
\subsection{Classical Turn-Arounds}
The classical turn-around, characterized by $\mubar c = m \pi$ and $\rho =0$,
is not a feature limited to LQC but would occur even in the classical theory. 
For standard perfect fluids with fixed equation of state this does not occur 
as the energy density is always non-zero. However, if the scalar field  potential is negative 
 or a  phantom field is present this condition can be met.
At this turn-around $\cos(\mubar c) =1$, $\sin(\mubar c) = 0$ which gives us
\begin{eqnarray}
        \ddot{p}\, |_{\dot{p}=0} &=& \frac{2}{3} \kappa p^{1/2} \frac{\partial \hm}{\partial
p} \nonumber \\
        &=& \kappa p^{-1/2} \left( \mp \half p^{-3/2} \Pp^2 + p^{3/2} V \right)
\end{eqnarray}
and we can again use $\rho = 0$ to relate $\mp \half p^{-3/2} \Pp^2 = p^{3/2} V$
giving
\begin{equation}
        \ddot{p}\, |_{\dot{p}=0}= 2 \kappa p V \,.
\end{equation}
                                                                                
In the case of a normal scalar field the potential must be negative in order
for the 
energy density to become zero hence $\ddot{p} < 0$ at turn-around and  a
recollapse
is possible. In the case of a phantom model we require the potential must be
positive and
hence a classical bounce can occur. 
Furthermore, in the course of dynamics it is possible
that both a quantum and classical turn-around can occur. This is 
precisely what can happen when we consider negative potentials
as the universe cycles between classical recollapses and quantum
bounces.

\section{Qualitative analysis of bounce with Inflationary potentials}
The analysis of the previous section has established analytically the turn-around behavior
{\em if} the conditions for a quantum or classical turn-around are met.
However the analysis does not indicate whether the turn-around
is a  feature of the dynamics given generic initial conditions. To go further
we will use phase portraits to analyze the space of solutions to the 
dynamics. To illustrate the techniques and as a simple non-trivial
model we first consider an ordinary scalar field with $\phi^2$ and $\phi^4$ potentials
conventionally used for chaotic inflationary scenarios. In the previous section
we showed that only a quantum bounce is possible. We are particularly interested
in the question if  inflation
 in the post bounce expanding universe is an attractor
in the space of solutions.

It is useful to first obtain 
$\dot H$ equation using Friedmann equation (\ref{FE1}) and the conservation law (\ref{EC}):
\begin{eqnarray}
\label{FE2}
	\dot H=-\frac{\kappa}{2}\left(\rho+p_\phi\right)\left(1-\frac{2\rho}{\rcr}\right)
\end{eqnarray}
where $\rho = \rho_\phi$ (\ref{edp}).

Already from (\ref{FE1}) and (\ref{EC}) it is evident that both energy density and Hubble parameter are bounded
\begin{eqnarray}
\label{Hb}
	-H_{max}\leq H \leq H_{max}=\sqrt{\frac{\kappa\rcr}{12}}, \\
\label{rhob}
	0 \leq \rho \leq \rcr, ~~~~~~~~\\
	\rho(H=\pm H_{max})=\frac{1}{2}\rcr, \\
	H(\rho=\rcr)=H(\rho=0)=0.
\end{eqnarray}
For the scalar field with non-negative  potentials, the only nontrivial case is a quantum turn-around, $H(\rcr)=0$ and it always represents a bounce, since $\rho+p_\phi=\dot\phi^2\geq 0$ and $\dot H>0$ in (\ref{FE2}). Therefore, given any positive definite potentials all solutions of the cosmological equations (\ref{EC}) and (\ref{FE1}) are {\it non-singular}.
As an example, let us discuss the  
simplest case of a  massless scalar field.
The solution of (\ref{EC}) can be written as
\begin{eqnarray}
	\dot\phi=\sqrt{2 \rcr} \left(\frac{a_c}{a}\right)^{3}.
\end{eqnarray}
Substituting this solution in Eq. (\ref{FE1}) gives the analytic solution for Hubble parameter
\begin{eqnarray}
	H=\pm \sqrt{2 \rcr} \frac{a_c^3}{a^6}\sqrt{\frac{\kappa}{6}(a^6-a_c^6)}.
\end{eqnarray}
Clearly this solution is devoid of any singularity as soon before and after the bounce the Hubble parameter reaches its maximum value.

When the effective potential of the scalar field does not vanish analytical solutions are difficult to find. To understand the qualitative behavior of solutions it is useful to build phase portraits.
The dynamical phase space is four dimensional (consisting of $c(t), p(t), \phi(t), \dot\phi(t)$)
however the vanishing of the constraint can be used to fix one of them
given the other three. Therefore a complete phase portrait
would be three-dimensional, however for simplicity
we will display two dimensional portraits consisting of
 $\phi$ and $\dot\phi$. Since the Hubble parameter only changes
sign at a classical or quantum turn-around each phase portrait
drawn will consist of either an expanding or contracting phase.
The quantum and classical turn-arounds will be represented
as boundaries in the phase portrait (which we will represent as solid lines
in the figures)
and any trajectory that reaches the boundary will continue
into the corresponding phase portrait with opposite sign of the Hubble rate
thus indicating the turn-around. In this sense the expanding and contracting
phase portraits are to be glued together at the boundaries.

%

In order to obtain equations for corresponding dynamical system
we can express the Hubble parameter in
terms of $\phi$ and $\dot \phi$ using the effective Friedmann
equation and plug into the conservation law Eq. (\ref{EC})
to get
\begin{equation}
\begin{array}{l}
	\ddot\phi=-\frac{\partial V}{\partial\phi}\mp 3\dot\phi\left\{\frac{\kappa}{3}\left(\frac{\dot\phi^2}{2}+V\right)\left[1-\frac{1}{\rcr}\left(\frac{\dot\phi^2}{2}+V\right)\right]\right\}^{\frac{1}{2}},
\end{array}
\end{equation}
where the sign ``--" corresponds to expansion while the sign ``+" gives contraction.
Using this equation we can obtain the phase space portraits
\cite{fn-p}  for the massive scalar field
$V(\phi)=m^2\phi^2/2$ with $m=0.2$ and self-interacting nonlinear scalar field with $V(\phi)=\lambda\phi^4/4$ with $\lambda=10^{-2}$.
As we have discussed in the previous section, in this case only quantum bounce is possible which occurs when $\rho =\rcr$, that is
\begin{eqnarray}
	\frac{1}{2}\dot\phi^2+V(\phi)=\rcr.
	\label{outerb}
\end{eqnarray}

Regions close to the outer boundary correspond to high energy density
with quantum effects dominant whereas the low energy limit
occurs near the origin in the $\phi,\dot\phi$ plane.

\begin{figure}[ht]
    \centering
        \includegraphics[width=3.5in]{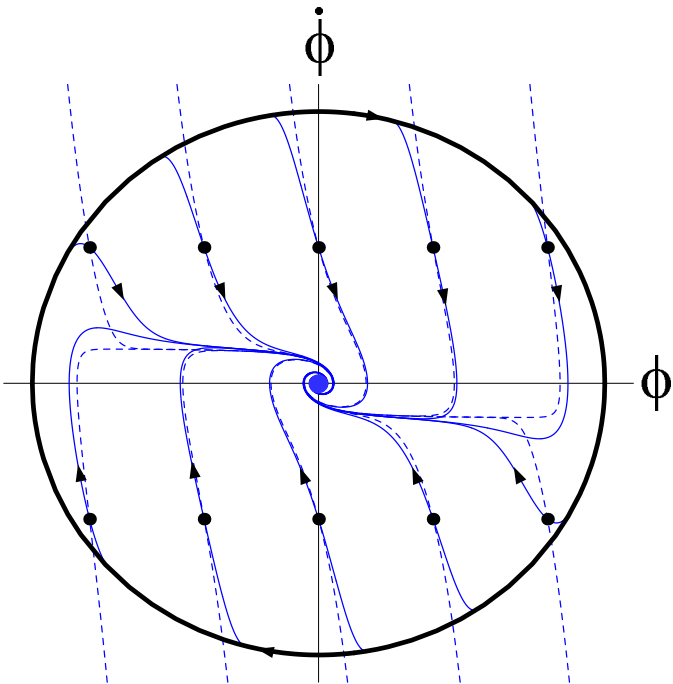}
    \caption{Phase portrait for massive scalar field $\phi^2$ potential. Dashed curves represent GR case and solid curves shown LQC case. Dots correspond to points where initial conditions for numerical solutions were given. The direction of phase trajectories is from the boundary to the origin. Here $m=0.2$.}
    \label{ph1}
\end{figure}
\begin{figure}[ht]
    \centering
        \includegraphics[width=3.5in]{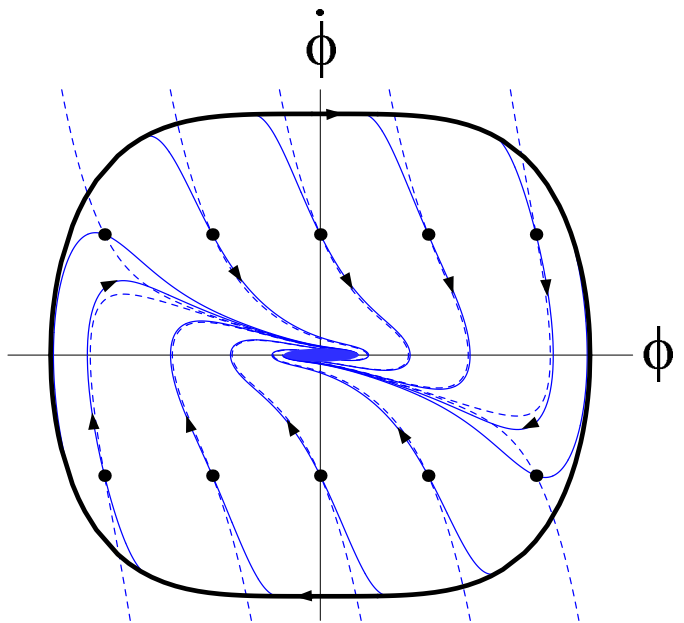}
    \caption{Phase portrait for the self-interacting scalar field  $\phi^4$ potential. Trajectories for LQC are shown by solid curves where as dashed curves show GR case.  Here $\lambda=10^{-2}$.}
%
    \label{ph2}
\end{figure}
The results for the two potentials in the expanding phase are
shown in 
fig. \ref{ph1} and fig. \ref{ph2} respectively.
The phase space trajectories obtained via LQC dynamics are compared with those of GR obtained in Ref. \cite{Belinsky1985} which are given as the dotted lines
in the figures. 
All phase trajectories are directed towards the origin which is the only particular 
point in the finite region of phase space having the character of a stable node. 
The scalar field experiences numerous damped oscillation around the
origin in agreement with classical GR. The agreement between
the two theories  is indicative  of the fact that
$\rho \ll \rcr$ near the origin.
In the region near
the outer boundary the LQC trajectories deviate radically from
the GR case and the existence of the boundary is a fundamental difference between   LQC and   GR \cite{bel-footnote}  (see analogous discussion for gauge theories of gravity case in \cite{Vereshchagin2003,Vereshchagin2005}).


A key feature of these phase portraits is the inflationary separatrix (see details for e.g. in Refs. \cite{Belinsky1985,bike,Vereshchagin2005}) which 
appears in the figures as the  curve in the second and fourth
quadrants to which the trajectories are attracted before undergoing
oscillatory behavior. It is here that inflationary behavior occurs and it was shown in Ref. \cite{Belinsky1985} that most solutions for $\phi^2$  and $\phi^4$ potentials within GR tend to this separatrix in the expansion stage. This led the authors of Ref. \cite{Belinsky1985} to the conclusion that inflation is generic within these models.
This separatrix is also present in the LQC case and is qualitatively the same although its position slightly changes with respect to the case of GR especially in the vicinity of the boundary.

\begin{figure}[hb]
    \centering
        \includegraphics[width=3.5in]{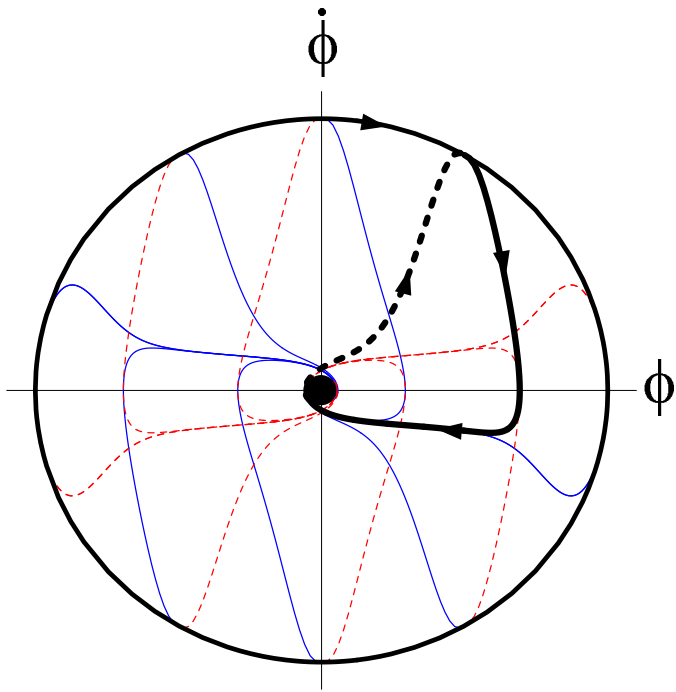}
    \caption{Complete phase portrait for massive scalar field within LQC showing both contracting (dashed curve) and expanding phases (full curve). Thick curve shows a concrete example of the phase trajectory.}
    \label{ph3}
\end{figure}

However, in LQC the trajectories do not begin at the outer boundary but
are preceded by a contracting phase before the bounce. 
The phase portraits can be obtained for the contracting phase
by reflecting the expanding portraits with respect to the vertical axis
and inverting the direction  of the trajectories.
Now all cosmological solutions start in the unstable node 
at the origin where the universe is contracting and the scalar field
is behaving as a driven oscillator  (anti-friction like behavior),
reach the outer boundary giving a bounce, return to the inflationary
separatrix in the expanding phase and finally approach the stable
node at the origin where the scalar field experiences damped oscillations.
We show the complete phase portrait of massive scalar field within LQC in fig.\ref{ph3} 
by superimposing the contracting and expanding phases.
The dashed curves correspond to the contracting phase and full curves  describe the expanding phase. 
The thick curve gives a concrete example of the closed phase trajectory which starts at the origin, comes to the boundary, experiences a non-singular bounce and comes back to the origin.


Due to the fact that inflationary separatrix is also present in the case of LQC we come to the conclusion that inflation is as common feature of cosmological solutions within LQC as it is within GR. 
If one  defines a measure of solutions on the boundary (at the bounce) it turns out that only a small fraction of solutions does not contain inflationary phase. 
Since all solutions are non-singular, they continue in the past to the contracting phase. The corresponding separatrix has a repulsive nature in contracting phase (i.e. solutions tend to deviate more and more from it in course of cosmological contraction), and the probability for solutions to follow this seperatix, given equal measure of initial conditions at the bounce, is small \cite{seperatrix-footnote}.
In fact this separatrix was shown to be exponentially unstable within GR, meaning that solution which is initially close to it deviates exponentially from it in course of time \cite{Vereshchagin2004}. Therefore, if all initial conditions have equal probability at the boundary, the probability to get non-inflating solution is exponentially suppressed. An example of the phase trajectory in fig.\ref{ph3} shows that solution can not follow repulsive separatrix on contracting stage but nevertheless has inflationary stage during expansion.

\section{Qualitative analysis of bounce with negative potentials}

Negative potentials have been used in Cyclic models to construct alternatives to inflationary scenarios. It is envisioned that seeds for the generation of cosmic structure originated in the contracting phase of the universe which preceded the current expanding phase. A basic difficulty in these scenarios is the singular nature of transition between two phases which has been extensively studied in Ref. \cite{bike}.  As we noted in Sec.II, a classical bounce is not possible 
with negative potentials for a normal scalar field, whereas a bounce mediated by loop quantum corrections is possible.  We now analyze the nature of this bounce for different negative potentials. 

\subsection{Massive scalar field}
For simplicity we start with massive scalar field with a constant negative potential
of form
\begin{eqnarray}
	V=\frac{m^2\phi^2}{2}-V_0,
\label{massneg}
\end{eqnarray}
where $V_0>0$. Without loss of generality
for the numerical simulations we use $V_0=0.1$ and $m=0.2$ in this subsection. For this potential there exists an inner and an outer boundary in phase portraits. The inner boundary arises due to classical recollapse, a feature shared with  classical GR \cite{bike}. The outer boundary 
represents the quantum bounce and is absent classically. The presence of the two boundaries  in LQC opens a novel 
possibility to construct cyclic models  where the universe undergoes a series of expansion and contraction phases.

\begin{figure}[htp]
    \centering
        \includegraphics[width=3.5in]{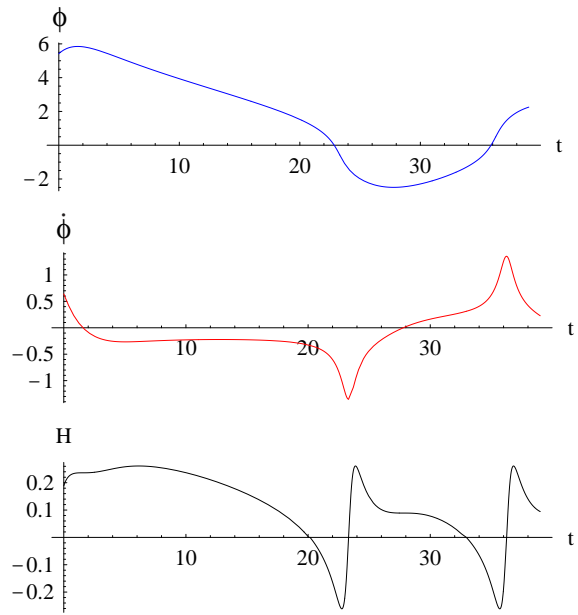}
    \caption{Behavior of $\phi$, $\dot \phi$ and $H$ is shown for the massive scalar field potential (\ref{massneg}) in LQC. The field starts rolling down the negative potential from positive value of $\phi$ and turns around due to loop quantum effects reaching the value where it started from. The scale factor also bounces as is evident from the turn-around in the sign of Hubble parameter. Initial conditions are $\dot\phi=0.640$, $\phi=5.422$; parameters are $V_0=0.1$, $m=0.2$.}
    \label{neg1a}
\end{figure}

\begin{figure}[hb]
    \centering
        \includegraphics[width=3.5in]{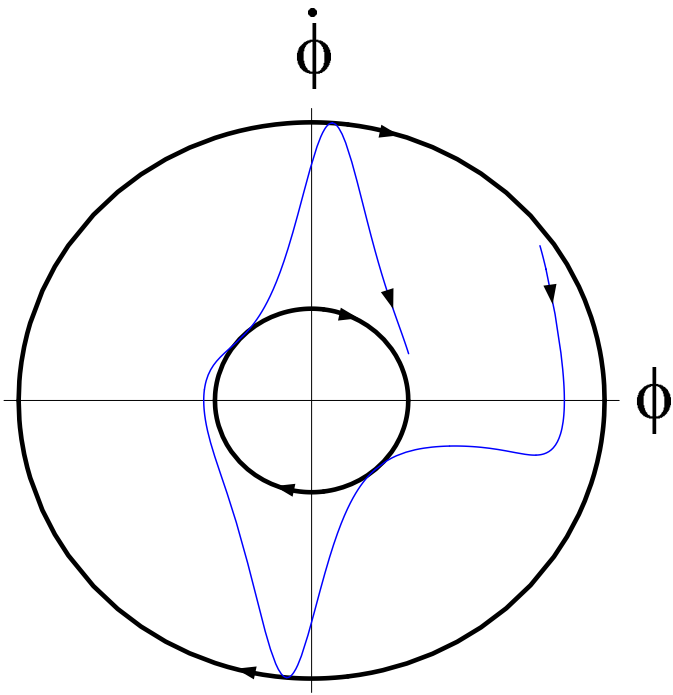}
    \caption{Phase trajectory of the solution shown in
fig.\ref{neg1a}. The inner boundary corresponds to classical bounce
($\rho = 0$) and the outer boundary to quantum bounce ($\rho =
\rcr$). The trajectory formally corresponds to infinite asymmetric oscillations, although is shown for a finite time interval.}
    \label{neg1b}
\end{figure}

Numerical solutions  for the time evolution given a set of initial conditions are shown in fig.\ref{neg1a} and the phase portrait is shown in fig.\ref{neg1b}. Both the Hubble parameter and energy density are bounded subject to the constraints given in (\ref{Hb}) and (\ref{rhob}). As can be seen from fig.\ref{neg1a}, the 
scalar field starts from  positive values and rolls down the potential
giving rise to a period of inflation with the Hubble parameter positive and
almost constant. This continues until the scalar field enters the region
of negative potential 
and the energy density reaches zero at which point the classical recollapse
occurs and  $H = 0$. In classical GR the field would continue its motion towards negative $\phi$ with ever increasing $\dot \phi$ (because of the anti-friction term 
in the Klein-Gordon equation) and finally ending in a big crunch singularity.
However, in LQC when $\dot \phi$ reaches a value such that the energy density of the field becomes comparable to $\rcr$, the trajectory deviates from the classical one and 
the magnitude of the Hubble parameter starts decreasing and quickly becomes zero at $\rho = \rcr$.  The universe bounces and immediately afterward 
enters a phase of super-inflation, where $\dot H >0$ (which has been shown to be a generic feature in LQC for $\rho > \rcr/2$ \cite{singh:2006a}). Since the universe expands quickly after the bounce, $\dot \phi$ decreases 
and $\rho$ becomes smaller than $\rcr/2$. The phase of super-inflation ends and Hubble rate starts decreasing. The field is now negative valued and climbs up the potential
only to turn-around, undergo slow-roll inflation and repeat the process 
in the opposite direction. The cycling continues indefinitely which is evident
from the phase portrait in fig.\ref{neg1b}.

As can be seen from fig.\ref{neg1a}, the oscillations of the dynamical variables are asymmetric. This is due to the fact that during expansion the inflationary separatrix (which is still present, like in the cases discussed in Sec.III) has an attractive nature, while during contraction it is repulsive one. Therefore the expanding universe spends more
time in the inflationary phase as opposed to the time spent in the contracting phase and the
universe experiences overall net growth in the scale factor.
The oscillations of the scalar field together with a monotonic Hubble parameter, which are typical for usual chaotic inflationary potentials, occur only if the inner boundary is close enough to the origin (corresponding to a small value of $V_0$) allowing
for the scalar field to oscillate before being reflected by the inner boundary.

As is evident from the phase portrait in fig.\ref{neg1b} given the
parameters chosen these oscillations
do not occur. The presence of the two boundaries leads to the only one global direction of phase trajectories, the clockwise direction seen from fig.\ref{neg1b}.
\begin{figure}[hb]
    \centering
        \includegraphics[width=3.2in]{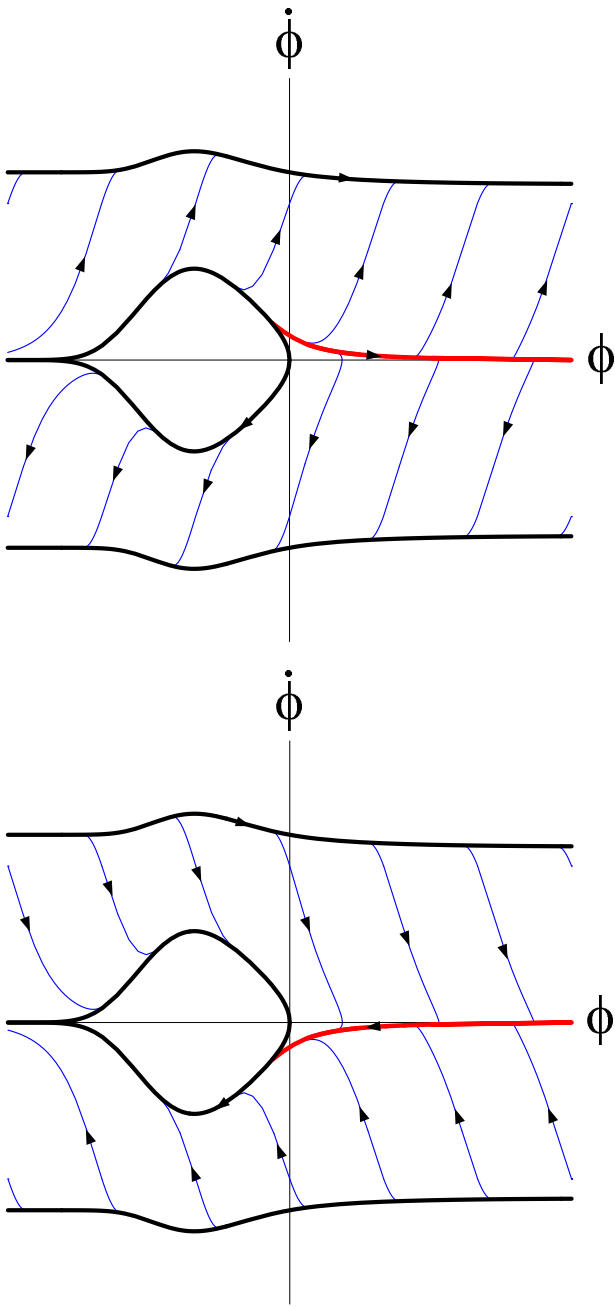}
    \caption{Phase diagrams for contraction (above) and expansion
(below) stages with the cyclic potential (\ref{cyclicp}). Red curve
shows repulsive (attractive) inflationary separatrix. Here
$\sigma=0.3\sqrt{8\pi}$, $\omega=0.09\sqrt{8\pi}$, and
$V_0=0.1$. Outer boundary corresponds to $\rho = \rcr$ and inner
boundary to $\rho = 0$.
}
    \label{cyc}
\end{figure}

\subsection{Cyclic potential}

Let us now consider the cyclic potential in the form \cite{StTur}
\begin{eqnarray}
V=V_0(1-e^{-\sigma \phi})\exp(-e^{-\omega \phi}).
\label{cyclicp}
\end{eqnarray}
In the Cyclic model,  the value $V_0$ is chosen so as to give
the correct magnitude of the current cosmic acceleration and the other
parameters are typically constrained to give the correct amplitude for
density perturbations. For the purposes of qualitative analysis and without any loss of generality, we choose the following parameters: $\sigma=0.3\sqrt{8\pi}$, $\omega=0.09\sqrt{8\pi}$, and $V_0=0.1$.
The minimum of the potential is displaced from the origin and the inner boundary is situated to the left of the origin. The phase diagrams for expansion and contraction stages with this potential are represented in fig.\ref{cyc}.

Here, as in the previous case, both a classical recollapse and quantum bounce
can occur. Again, there is a separatrix in the phase space located to the right from the inner boundary. As in the previous cases during contraction phase trajectory is directed towards the outer boundary and during expansion it is directed towards the inner boundary.

\begin{figure}[htp]
	\centering
		\includegraphics[width=3.5in]{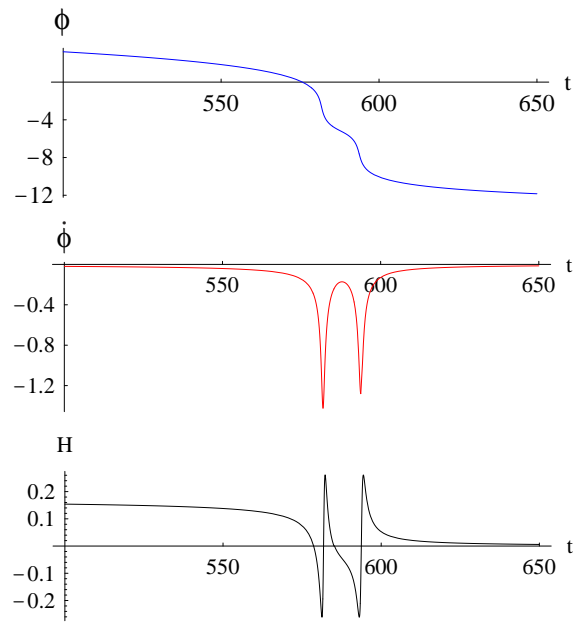}
	\caption{Behavior of $\phi$, $\dot \phi$ and $H$ with the
Cyclic model potential (\ref{cyclicp}). Initial conditions for this solution are: $\dot\phi=16/3$, $\phi=1.066$. Parameters are: $\sigma=0.3\sqrt{8\pi}$, $\omega=0.09\sqrt{8\pi}$, and $V_0=0.1$}
	\label{cyc_t}
\end{figure}

The evolution of the Hubble parameter in LQC with the Cyclic model potential can be understood from fig. \ref{cyc_t}. As in the previous case we investigate the dynamics when the field starts from positive $\phi$ where the potential is positive. The field rolls down towards negative values of $\phi$ and upon entering the region of negative potential  its kinetic energy cancels the potential energy 
leading to $H = 0$ and the classical recollapse. The universe starts contracting, heading towards a big crunch but experiences a quantum bounce when 
$\rho = \rcr$. After the bounce, the universe briefly enters the phase of super-inflation before the Hubble parameter starts decreasing. However, unlike the previous case
the potential is still negative valued and the field can not turn around and
continues to $\phi = - \infty$. Another round of collapse and bounce occurs,
but this behavior terminates and the universe approaches the state
$H \rightarrow 0$ with $\dot \phi \rightarrow 0$ and $V \rightarrow 0$ at $t \rightarrow \infty$. Thus in the case of cyclic model potential  
though the scale factor bounces and universe escapes the big crunch singularity, the dynamics does not lead to cycles. For such cycles to exist the scalar field must
turn around and return to the region of positive $\phi$. To achieve
this the potential must become positive at some point for negative $\phi$ as was
the case for the quadratic potential with negative constant. Since the Cyclic model potential is negative for all values of $\phi < 0$ and approaches zero asymptotically, the possibility of such a turn-around in $\phi$ does not exist.

\begin{figure}[htp]
	\centering
		\includegraphics[width=3.5in]{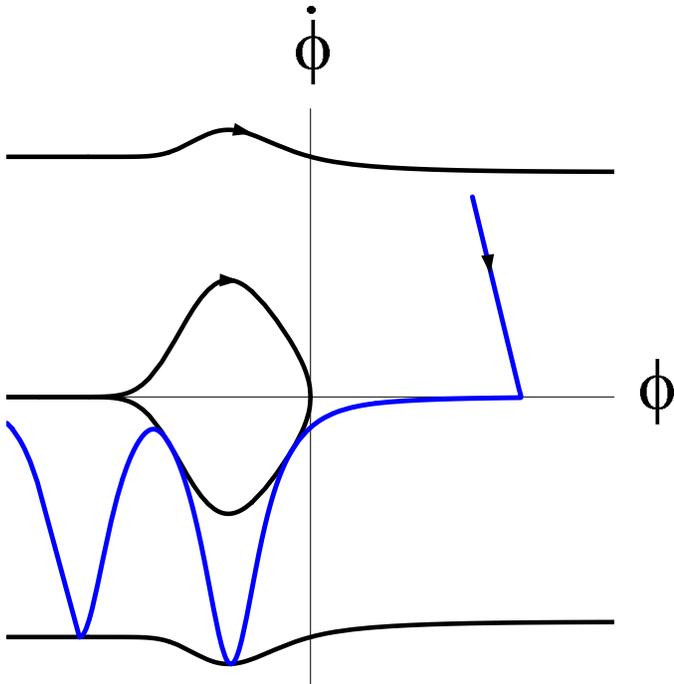}
	\caption{Phase portrait of solution shown at fig.\ref{cyc_t},
for wider time interval. Here again we draw only finite time evolution
of the solution. Inner boundary corresponds to $\rho = 0$ and outer to
$\rho = \rcr$.}
	\label{cyc_tr}
\end{figure}

The structure of the phase space shown at fig.\ref{cyc} suggests that all solutions begin with $\phi=-\infty$ and $\dot\phi>0$ and end at $\phi=-\infty$ with $\dot\phi<0$. The boundaries define the overall direction for the phase space flow which is clockwise, like in the case of massive negative potential, but the outer boundary is not closed, so the scalar field  escapes to negative infinity. This is illustrated at fig.\ref{cyc_tr} with a particular solution. We have set initial conditions in a way to have positive $\phi$ at the beginning with the purpose to illustrate the dynamics with cyclic potential within LQC. However the evolution can be traced continuously to the past with the field having large negative values.

The problem of non-singular bounces for Cyclic potential in LQC has been investigated earlier \cite{cyclic}, but 
by considering modifications to only matter part of the Hamiltonian. The  analysis of Ref. \cite{cyclic} 
was able to show the existence of non-singular bounce in scale factor
for Cyclic model for some choices of parameters but not in general. A
limitation of that analysis was the problem of energy density becoming
super-Planckian as the field rolls down the Cyclic potential
\cite{cyclic1}. As we understand from the present analysis, incorporation of modifications to the gravitational part in the effective Hamiltonian automatically solve this problem, as the energy density is bounded 
above by a critical value.

\subsection{Bicyclic potential}

The Bicyclic potential was discussed in Refs. \cite{bike,Linde} as a better 
choice over the Cyclic potential in the attempt to construct cosmologically viable 
cyclic models. In fact there exists the crucial difference between the
cyclic and inflationary models. The principle difficulty of the former
is the cosmic singularity, but in a much more drastic way than appears
in standard inflation. In contrast to inflation, within cyclic models
perturbations are created at the contraction stage and cannot be
traced through singularity. Instead, in our case, given regular
background dynamics  perturbations can in principle be evolved through the
regular bounce once these are incorporated in LQC.
At present we are  interested in comparing its phase portraits with the Cyclic potential. 

\begin{figure}[htp]
    \centering
        \includegraphics[width=3.5in]{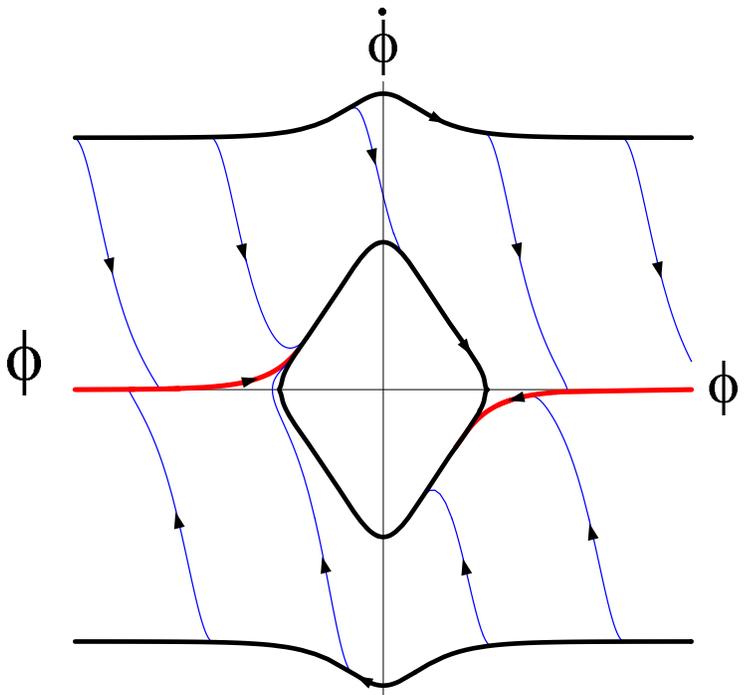}
    \caption{Phase diagram for expansion stage with the Bicyclic
potential (\ref{bicyclic}). Red curve again shows attractive
inflationary separatrix. Inner boundary refers to $\rho = 0$ and outer
to $\rho = \rcr$. Here $A=10$ and $V_0=0.03$.}
    \label{cyc1}
\end{figure}

The form of the Bicyclic potential we will discuss is
\begin{eqnarray}
	V=V_0\left[1-A\cosh^{-1}(\phi)\right],
\label{bicyclic}
\end{eqnarray}
where $A>0$ and $V_0>0$. For qualitative analysis we choose $A=10$ and $V_0=0.03$. The phase diagram for the expansion stage is represented in fig.\ref{cyc1}. As in the previous cases there exist inner and outer boundaries corresponding
to classical recollapses and quantum bounces. As can be seen from the phase portrait the outer boundary becomes a horizontal line far from the origin given by the condition $\dot\phi_b=\pm\sqrt{2\rcr}$ since for large values of $\phi$ the potential is small.
\begin{figure}[htp]
    \centering
        \includegraphics[width=3.5in]{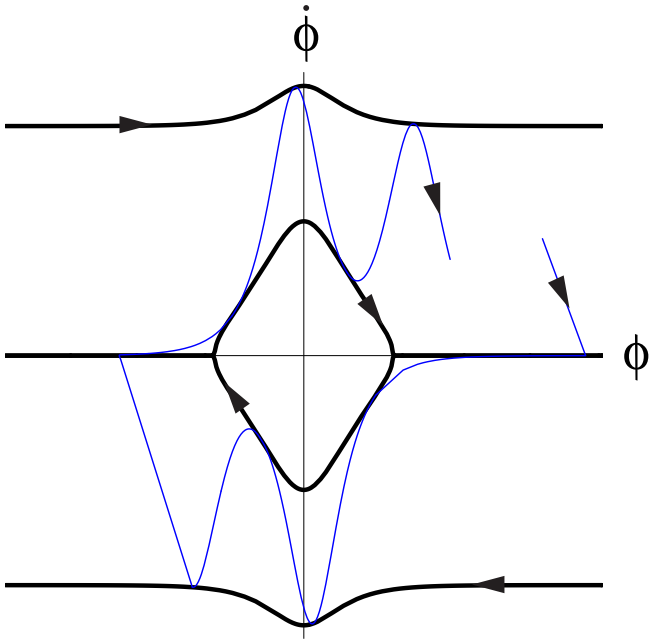}
    \caption{Phase portrait with the Bicyclic potential
(\ref{bicyclic}) for a finite time interval. Outer boundary
corresponds to $\rho = \rcr$ and inner boundary refers to $\rho = 0$.}
    \label{bicyc}
\end{figure}
The inner boundary is a closed curve surrounding the origin crossing the $\phi$ axis at points where the potential (\ref{bicyclic}) vanishes, namely when $\phi=\pm\textrm{arccosh}(A)$.

The phase portraits for the contracting branch can be obtained by reflecting the phase diagram in fig.\ref{cyc1} with respect to the vertical axis and inverting the directions of phase trajectories. Example of phase trajectory for bicyclic potential is represented at fig.\ref{bicyc}.

A difference between the Bicyclic model and the previous  models  is that there is  a 
closed inner boundary but an open outer boundary. However as  in the case of massive scalar field with negative constant there is a preferred clockwise direction for phase trajectories.
In contrast to the Cyclic model, the behavior cycles both
in terms of the Hubble expansion and the scalar field.
Given initial conditions between the boundaries, the solution is attracted to the inflationary separatrix until reaching the inner boundary. After recollapsing
the separatrix is repulsive and the solution is driven to the
outer boundary leading to a quantum bounce and the process starts anew. Thus, with the Bicyclic potential there is an infinite number of inflationary stages within a given solution. The time variation of $\phi$, $\dot \phi$ and $H$ is qualitatively similar to that for the massive scalar field shown in fig.\ref{neg1a} with a bounce in both the scalar field and the scale factor. The Bicyclic model
thus leads to  a non-singular cyclic model in LQC. 
Again the crucial distinction between the Cyclic and Bicyclic potential
is the fact that the potential is positive for both large negative and positive
values of $\phi$. This cycling in the Bicyclic model is responsible
for the better viability of the latter.


\section{Discussion and conclusions}

The non-perturbative quantum geometric effects lead to a $\rho^2$ modification with a negative sign in the Friedmann equation in LQC. Since the correction term is negative definite it can lead to a quantum bounce in the high energy regime when loop quantum modifications are dominant. We have investigated the qualitative details of this bounce for inflationary and negative potentials. The existence of the outer boundary given by (\ref{outerb}) in the phase portraits guarantees non-singular behavior of all the solutions with a scalar field in LQC with any kind of potential. For negative potentials the inner boundary also appears corresponding
to the classical recollapse. The presence of the two boundaries for negative potentials leads to the possibility of solutions having cyclic behavior. We have also briefly reported on the 
nature of quantum turn-arounds for a phantom field.

The massless scalar field gives a good example of one feature of cosmological solutions within LQC, namely the non-singular bounce. In effect, at least solutions with $\phi^2$ and $\phi^4$ potentials can be well approximated near the bounce by the corresponding exact solutions for a  massless scalar field, because most of them approach the outer boundary of phase space with kinetic energy  much larger than the potential energy, thus the potential can be neglected. For these potentials we showed that inflationary trajectories are attractors of dynamics after the bounce.
In the same way a massive scalar field with negative potential represents a good example of cycling between expansion and contraction accompanied with inflation, 
exhibiting the main features of the more complicated potentials.
Our analysis shows that there is little qualitative difference between the massive scalar field with negative constant and Bicyclic potentials, although their functional form is very different. In both cases cyclic solutions can be found with an infinite number of inflationary stages. 

Though the problem of the big crunch can be overcome with potential of the Cyclic model, a limitation remains in the lack of a  turn-around of $\phi$ from the negative side of the potential.
 In the Cyclic model this instant is envisioned as the collision of two branes. If the effective potential between branes can become positive prior to collision (arising
from higher order perturbative string effects/or non-perturbative corrections), then a turn-around of the field can occur and a viable cyclic model can be constructed. On the other hand, the Bicyclic potential provides the possibility for the field to turn around and may be used for further development of cyclic models.

The qualitative analysis reported in this work has shown that both inflationary paradigm and the Cyclic/Bi-cyclic 
scenarios may be incorporated in LQC. Our analysis has not touched on the aspects which can 
quantify the viable parameter space of these models with respect to
observational data. Further, we have not explored the way LQC may 
shed insights on the problem of evolution of perturbations through the
bounce. 
Work on including perturbations in the quantum theory and
deriving effective equations is in progress, which will eventually address the viability of these ideas with 
respect to observations.

\acknowledgments{We thank V.A. Belinsky, Roy Maartens and Shinji Tsujikawa for useful discussions. Work of PS and KV is supported by NSF grants 
PHY-0354932 and PHY-0456913 and the Eberly research funds of Penn State.}


\begin{thebibliography}{10}
\bibitem{eff_ham_fn} Additionally the effective Hamiltonian may contain modifications due to
the inverse scale factor operator both in the gravitational \cite{kevin_ham}
and the matter part \cite{mblr,freq} which arise by expressing
inverse powers of the triad in terms of Poisson brackets between
holonomies and positive powers of the triad. These modifications
become important at a scale determined by a half-integer parameter $j$
in LQC. This scale is independent of the critical energy density (\ref{rcr}) which arises in Eq.(\ref{FE1}). For semi-classical states, if the value of $j$ is
small, which theoretical considerations indicate \cite{alex,kevin_ham},  effects due to inverse scale factor modifications become
weak in comparison to $\rho^2$ modifications (Eq.(\ref{FE1})) and can
be neglected. Due to this reason the modified Friedmann equation in
this work is different from the early works in LQC (see for eg. Ref. \cite{mblr}).
\bibitem{fn-p} Following numerical analysis is performed in Planckian
units.
\bibitem{seperatrix-footnote} Separatrix is not a singular curve. Given phase trajectory can approach it very closely but never joins it.
\bibitem{cyclic1} This is a well known problem with the potential in Cyclic model \cite{Linde}. In LQC treatment of negative potentials this issue has also been discussed in Ref. \cite{qmw}.
\end{thebibliography}
\end{document}